\documentclass[twocolumn,prb,aps,superscriptaddress,showpacs]{revtex4}
\usepackage{mathrsfs}
\usepackage{graphicx}
\begin{document}
\title {Dynamical revivals of the paired superfluidity and counterflow superfluidity}
\author{Shi-Jie Yang\footnote{Corresponding author: yangshijie@tsinghua.org.cn}}
\affiliation{Department of Physics, Beijing Normal University, Beijing 100875, China}
\affiliation{State Key Laboratory of Theoretical Physics, Institute of Theoretical Physics, Chinese Academy of Sciences, Beijing 100190, China}
\author{Pei Lu}
\author{Shiping Feng}
\affiliation{Department of Physics, Beijing Normal University, Beijing 100875, China}
\begin{abstract}
The quantum dynamics of two-species bosons in an optical lattice is studied within the mean-field theory. The quantum coherence experiences periodical collapses and revivals, which depends on the relative strength of the inter- and intra-species interactions. The paired superfluidity and the counterflow superfluidity are identified and are verified by exact diagonaliztion in the two-site Hubbard model. We suggest a dynamical method of implementation and detection for the paired superfluid and the counterflow superfluidity in the optical lattices.
\end{abstract}
\pacs{67.85.Fg, 67.85.De, 03.75.Kk, 03.75.Gg} \maketitle

Ultracold atoms in the optical lattices have been providing versatilely and experimentally feasible means to explore various fundamental problems in condensed matter physics\cite{Paganelli,Xu,Albus,Imambekov,Messeguer,Chatterjee,Zhou,Tsuchiya,Sachdev,Javanainen}. The external potentials, the dimensions, the effective interactions, as well as the atomic components can be well-controlled and precisely measured. The superfluid-Mott transition (MIT) is one of the most prominent phenomena. Other phenomena such as supersolid, atomic Josephson oscillations and macroscopic self-trapping, quantum tunneling, quantum coherence, disorder and impurity effects, are also extensively explored\cite{Anker,Albiez,Winkler,Will,Yang,Zhang,Pei,Lee1,Lee2}.

The situations become more subtle when a mixture of two-species bosons are presented in the lattices. In the limit of strong inter-particle interactions, the single particle tunneling is damped while the correlated co-tunneling of atom pair or anti-pair through the second quantum transition may dominate the motion of the system due to the constraint of energy conservation. The atoms may move across the lattice only by pairs or anti-pairs (particle-hole pairs), forming phases of the paired superfluid (PSF) or the counterflow superfluid (CFS) which are respectively characterized by the mean-field order parameters $\langle{\hat a}{\hat b}\rangle\neq 0$ or $\langle{\hat a^\dagger}{\hat b}\rangle\neq 0$ and $\langle{\hat a}\rangle=\langle{\hat b}\rangle=0$ in the meanwhile. Here ${\hat a}$ and ${\hat b}$ are the annihilation operators of the species-A and species-B bosons. These novel phases were predicted within the mean-field theory and verified by quantum Monte Carlo simulations or density-matrix renormalization group (DMRG) method\cite{Kuklov,Demler,Kagan,Svistunov,Yang1}. However, the experimental signals of the PSF and CFS are still absent.

Monitoring the non-equilibrium evolution of a multi-atom system serves as an alternative way to probe the dynamical behaviors and quantum coherence experimentally. The time evolution of the ensemble can be detected by analyzing the visibility of the atomic interference pattern after rapidly switching off the lattice potential and the subsequent time-of flight expansion. Greiner et al studied the dynamical evolution of the multiple matter wave interference pattern by adiabatically changing a superfluid (SF) into the deep Mott insulator (MI) regime\cite{Greiner}. The timescale for the jump in the potential depth is fast compared with the tunneling time between neighboring wells, but sufficiently slow to ensure that all atoms remain in the vibrational ground state of each well. They observed that the matter wave field of the Bose-Einstein condensate (BEC) undergoes a periodic series of collapses and revivals. This observation demonstrate behavior of ultracold matter beyond mean-field theories.

In this paper, we suggest that this technique may be employed to explore the dynamical evolution of PSF and the CFS, which will decidedly demonstrate the existence of such novel phases via experiments. We reveal the periodic collapses and revivals of the quantum coherence in each species bosons and between the two-species bosons by evaluating various mean-field theory order parameters. The revival periods are different and depend on the relative strength of the inter- and intra-species interactions. Most intriguingly, we observe that there exist time intervals that the complex order parameters $|\langle\hat{a}\hat{b}\rangle(\tau)|\neq 0$ or $|\langle\hat{a}^\dag\hat{b}\rangle(\tau)|\neq 0$ while $|\langle\hat{a}\rangle(\tau)|=|\langle\hat{b}\rangle(\tau)|=0$, indicating the periodic occurrence of PSF or CFS.

The mean-field theory findings are verified by numerical simulations with two-species bosons in a double well. Under the tight binding and single-orbit approximation, the double-well system is simplified to a two-site Bose-Hubbard model\cite{Jaksch,Vardi}. We compare the quantum coherence for different relative strength of the interactions between A-A bosons, B-B bosons, as well as A-B bosons by exact diagonalziation of the Hubbard model. The results coincide with the mean-field prediction quite well. Our results may provide a critical clue to determine the existence of PSF and CFS which can be directly observed in experiments.

The initial SF state is represented by a coherent state $|\psi\rangle=|\psi_A\rangle\times|\psi_B\rangle$ of the two-species (A and B) atomic matter fields. Such a coherent state can be expressed as a superposition of different number states $|n_A,n_B\rangle$ such that $|\psi_\sigma\rangle=\exp(-|\upsilon_\sigma|^2/2)\sum_{n_\sigma}\frac{\upsilon_\sigma^{n_\sigma}}{\sqrt{n_\sigma!}}|n_\sigma\rangle$, ($\sigma=A,B$), where $\upsilon_\sigma=\sqrt{\bar{n}_\sigma}e^{i\phi_\sigma}$ denote the complex field amplitude with mean bosonic number $\bar{n}_\sigma$ and initial phase $\phi_\sigma$. This wavefunction does not incorporate any information on the geometry or dimensionality of the lattice and is therefore mean-field in nature\cite{Krauth}.

As the system is swiftly tuned to the deep MI regime, it is in a superposition of different eigenstates which are Fock states in the atom number. The on-site eigenenergies are $E_{m,n}=\frac{1}{2}U_{AA}m(m-1)+\frac{1}{2}U_{BB}n(n-1)+U_{AB}mn$, where $U_{AA}$, $U_{BB}$ and $U_{AB}$ are the Hubbard interactions between the A-A, B-B and A-B bosons, respectively. For clearance, we will use $m$ and $n$ represent the atom number of species-A and species-B atoms, respectively. The state evolves in time according to their eigenstates as,
\begin{equation}
|\psi(\tau)\rangle=\sum_{m,n=0}^N c_{m}d_{n}e^{-iE_{m,n}\tau}|m\rangle|n\rangle.\label{state}
\end{equation}
The coefficients read $c_m=\frac{1}{\sqrt{m!}}e^{-|\upsilon_A|^2/2}\upsilon_A^m$ and $d_n=\frac{1}{\sqrt{n!}}e^{-|\upsilon_B|^2/2}\upsilon_B^n$, which satisfy the normalization conditions $\sum_{m}|c_m|^2=\sum_{n}|d_n|^2=1$. The interactions between the two species bosonic fields are encoded in the dynamics of the on-site wave function.
\begin{figure}
\begin{center}
\includegraphics*[width=8.5cm]{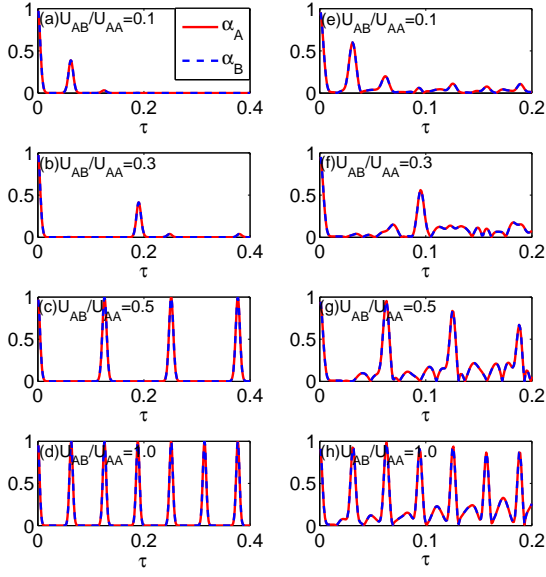}
\caption{(Color online) Collapses and revivals of the individual coherence of species-A (solid/red line) and species-B (dashed/blue) bosons for $U_{BB}/U_{AA}=1.0$. (a-d) The mean-field theory predictions for $U_{AB}/U_{AA}=0.1$, $0.3$, $0.5$, and $1.0$, respectively. (e-h) are the corresponding numerical results by the two-site Hubbard model. $U_{AA}=100$ and $t_A=t_B=1$.}
\end{center}
\end{figure}

The evolution of the macroscopic matter wave fields are evaluated by the average of atomic field operator $\hat a$ ($\hat b$) in the state (\ref{state}) which yield\cite{Will},
\begin{eqnarray}
\alpha_A(\tau)&\equiv&|\langle\hat{a}\rangle|/\sqrt{\bar{m}}=e^{-\bar{m}[1-\cos(U_{AA}\tau)]-\bar{n}[1-\cos(U_{AB}\tau)]},\nonumber\\
\alpha_B(\tau)&\equiv&|\langle\hat{b}\rangle|/\sqrt{\bar{n}}=e^{-\bar{n}[1-\cos(U_{BB}\tau)]-\bar{m}[1-\cos(U_{AB}\tau)]}.
\label{alpha}
\end{eqnarray}
The parameters $\alpha_A$ and $\alpha_B$ indicate the ratios of the atoms in the macroscopic matter waves (or the coherent state) over the mean total atoms in each lattice site and hence reflect the degree of coherence in the many-boson system.

\begin{figure}
\begin{center}
\includegraphics*[width=8.5cm]{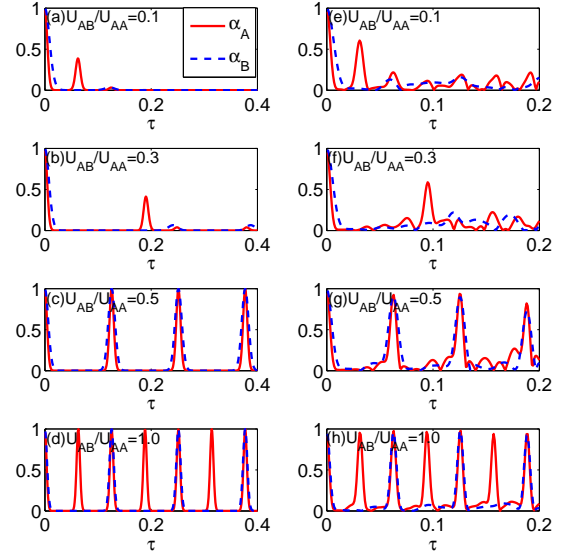}
\caption{(Color online) The same as in Fig.1 for $U_{BB}/U_{AA}=0.5$. The odd revival peaks are damped as $U_{AB}/U_{AA}=1.0$.}
\end{center}
\end{figure}

One observes that the evolution of the matter wave field of one species boson $\alpha_A$ ($\alpha_B$) is irrelevant to the intra-interaction of the other species bosons $U_{BB}$ ($U_{AA}$). Figure 1(a-d) compare the time evolution of each species for equal intra-specie interactions $U_{BB}=U_{AA}$ at various inter-species interactions $U_{AB}/U_{AA}=0.1,0.3,0.5,1.0$. The mean atom numbers are $\bar m=\bar n=5$. In these cases the two species evolve synchronously, $\alpha_A(\tau)=\alpha_B(\tau)$. As the intra-species interaction is commensurate with the inter-species interaction the coherence experience periodic collapses and revivals, with the revival period $T_A=T_B=2\pi/U_{AB}$ (Fig.1(c,d)). At the first, the interference of different phases of the atom number states lead to a collapse of $\alpha_A$ ($\alpha_B$). After integer multiples of the revival time $T_A$ ($T_B$), all phase factors in the sum of equation (\ref{state}) re-phase modulo $2\pi$, leading to revivals of the initial coherent state. On the other hand, if the intra- and inter-species interactions are incommensurate, the phase factors in Eq.(\ref{state}) cannot re-phase (modulo $2\pi$) and the matter wave fields gradually damped (Fig.1(a,b)).

When the intra-species interactions are different $U_{AA}\neq U_{BB}$, the evolution for each species exhibits distinct features. Figure 2(a-d) show the coherence evolution for $U_{BB}/U_{AA}=0.5$. While $\alpha_A(\tau)$ is almost the same as in Fig.1, the odd revival peaks are damped in $\alpha_B(\tau)$ (Fig.2(d)). Hence the revival period $T_B=2\pi/U_{BB}=2T_A$. The phenomena are somehow similar to the experimental observation of the damp of odd revival peaks by fermionic impurities\cite{Will,Pei}.

The mean-field results can be verified by exact diagonalization within the two-site Bose-Hubbard model. Consider two-species $N_A$ and $N_B$ bosons, the Hamiltonian reads\cite{Jaksch,Vardi}
\begin{eqnarray}
\hat{H}=&&-t_A(\hat{a}_1^\dag \hat{a}_2+\hat{a}_2^\dag
\hat{a}_1)-t_B(\hat{b}_1^\dag \hat{b}_2+\hat{b}_2^\dag
\hat{b}_1)\nonumber\\
&&+\frac{1}{2}U_{AA}\sum_{i=1,2}\hat{m}_{i}(\hat{m}_{i}-1)+\frac{1}{2}U_{BB}\sum_{i=1,2}\hat{n}_{i}(\hat{n}_{i}-1)\nonumber\\
&&+U_{AB}(\hat{m}_{1}\hat{n}_{1}+\hat{m}_{2}
\hat{n}_{2}),\label{ham1}
\end{eqnarray}
where $t_A$ and $t_B$ are the hopping coefficients, $\hat m$ and $\hat n$ are the number operators of species-A and species-B bosons, respectively. For convenience, we choose $t_A=t_B=1$ as the units.

The Hamiltonian (\ref{ham1}) is represented in the $(N_A+1)\times(N_B+1)$ Fock basis set $\{|0,N_A\rangle|0,N_B\rangle,|1,N_A-1\rangle|0,N_B\rangle,\cdots,|k,N_A-k\rangle|0,N_B\rangle,
\cdots,|k,N_A-k\rangle|l,N_B-l\rangle,\cdots,|N_A,0\rangle|N_B,0\rangle\}$. The eigenstates can be expressed as linear combinations of the Fock bases,
$|\psi_j\rangle=\sum_{k,l=0}^{N_A,N_B} c_{j,[k+l(N_A+1)]}|k,N_A-k\rangle|l,N_B-l\rangle$ ($j=0,1,2,\ldots,N$), which correspond to the
eigenvalues $\omega_j$. The coefficients $c_{j,[k+l(N_A+1)]}$ satisfy the recursive relations
\begin{widetext}
\begin{eqnarray}
&&-t_A\sqrt{(N_A-k)(k+1)} c_{j,[k+1+l(N_A+1)]}-t_A\sqrt{(N_A-k+1)k} c_{j,[k-1+l(N_A+1)]}\nonumber\\
&&-t_B\sqrt{(N_B-l)(l+1)} c_{j,[k+(l+1)(N_A+1)]}-t_B\sqrt{(N_B-l+1)l} c_{j,[k+(l-1)(N_A+1)]}\nonumber\\
&&+[\frac{1}{2}U_{AA}(N_A^2-2N_Ak-N_A+2k^2)+\frac{1}{2}U_{BB}(N_B^2-2N_Bl-N_B+2l^2)\nonumber\\
&&+U_{AB}(N_AN_B-N_Al-N_Bk+2kl)-\omega_j] c_{j,[k+l(N_A+1)]}=0.
\end{eqnarray}
\end{widetext}

Starting from an initial coherent state
\begin{equation}
|\psi(0)\rangle=\frac{1}{2^{(N_A+N_B)/2}}({\hat a}_1^\dag+{\hat a}_2^\dag)^{N_A}({\hat b}_1^\dag+{\hat b}_2^\dag)^{N_B}|0\rangle,\label{initial}
\end{equation}
we find that the state evolves in time according to the eigenstates as
\begin{eqnarray}
|\psi(\tau)\rangle&=&\sum_{j=0}^Nf_j(\tau)|\psi_j\rangle \nonumber\\
&=&\sum_{k,l=0}^{N_A,N_B} g_{k,l}|k,N_A-k\rangle|l,N_B-l\rangle,
\end{eqnarray}
where $f_j(0)=\langle\psi_j|\psi(0)\rangle$ and $g_{k,l}(\tau)=\sum_{j=0}^Nf_j(0)c_{j,[k+l(N_A+1)]}e^{-iw_j\tau}$.

The coherence degrees $\alpha_A$ and $\alpha_B$ of each species bosons are characterized by the non-diagonal elements of the single-particle density matrix $\rho_{\mu\nu}^{(1)}(\tau)=\langle\psi(\tau)|{\hat a}_\mu^\dagger{\hat a}_\nu|\psi(\tau)\rangle$ (with ${\hat a}\rightarrow {\hat b}$ for B-bosons)\cite{Yang,Qi},
\begin{equation}
\alpha_{A}(\tau)=\frac{|\langle\psi(\tau)|{\hat a}_1^\dag{\hat a}_2|\psi(\tau)\rangle|}{|\langle\psi(0)|{\hat a}_1^\dag{\hat a}_2|\psi(0)\rangle|}.
\end{equation}
For comparison, the coherence degrees have been normalized to those of the initial state. In the following, we take $N_A=N_B=10$ and $U_{AA}=100$ as examples. Figure 1(e-h) and Fig.2(e-h) display the exact numerical results of the Hubbard model. Obviously, they agree with the mean-field theory quite well. Note that the time scales differ a factor of two in the two methods. It origins from the fact that in the mean-field theory we take account of only one (two) operator(s) while in the double well model two (four) operators are involved.

\begin{figure}
\begin{center}
\includegraphics*[width=8.5cm]{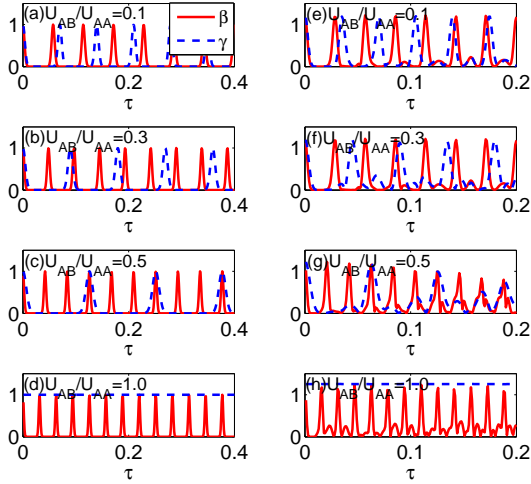}
\caption{(Color online) Periodic collapses and revivals of the heteronuclear paired coherence $\beta(\tau)$ (solid/red line) and the counterflow coherence $\gamma(\tau)$ (dashed/blue) for $U_{BB}/U_{AA}=1.0$. (a-d) The mean-field theory results. (e-h) are the corresponding numerical results by the two-site Hubbard model.}
\end{center}
\end{figure}

Based on the mean-field theory, we can evaluate the averages of the composite operators,
\begin{eqnarray}
\beta(\tau)&\equiv&|\langle\hat{a}\hat{b}\rangle|/\sqrt{\bar{m}\bar{n}}\nonumber\\
     &=&e^{-\bar{m}(1-\cos{(U_{AA}+U_{AB})\tau})-\bar{n}[1-\cos{(U_{BB}+U_{AB})\tau}]},\nonumber\\
\gamma(\tau)&\equiv&|\langle\hat{a}^\dag\hat{b}\rangle|/\sqrt{\bar{m}\bar{n}}\nonumber\\
     &=&e^{-\bar{m}[1-\cos(U_{AA}-U_{AB})\tau]-\bar{n}[1-\cos(U_{BB}-U_{AB})\tau]}.\label{beta}
\end{eqnarray}
The finite $\langle{\hat a}{\hat b}\rangle$ represents the correlation between the two-species bosons. It is a hetero-nuclear atom pair which implies strong correlations between the two species bosons. By comparing formula (\ref{alpha}) and (\ref{beta}), we find that the temporal evolution of $\beta(\tau)$ and $\gamma(\tau)$ exhibit different collapse and revival modes with $\alpha_{A,B}(\tau)$. $\beta(\tau)$ and $\gamma(\tau)$ depend on all of the three interactions while $\alpha(\tau)$ depends only on two of the interactions. It means that the revival periods of these physical parameters can be manipulated by properly adjust the interactions. The atom-pairs form PSF provided, in the meanwhile, that the single particle order parameter of each species vanishes, i.e., $\langle{\hat a}\rangle=\langle{\hat b}\rangle=0$\cite{Kuklov}. Similarly, the order parameter $\langle{\hat a}^\dagger{\hat b}\rangle$ reveals the correlation between the propagation of A-atoms and the counter-propagation of B-atoms. It is usually interpreted as the counterflow pair or anti-pair of a species-A particle with a species-B hole. The counterflow pairs form CFS whenever $\langle{\hat a}^\dagger{\hat b}\rangle\neq 0$ while $\langle{\hat a}\rangle=\langle{\hat b}\rangle=0$.

In the double well $\beta$ and $\gamma$ are respectively interpreted as coherence of a pair or anti-pair of particles tunneling through the barrier of the potential. Analogously, they are characterized by the non-diagonal elements of the two-particle density matrices as,
\begin{eqnarray}
\beta(\tau)&=&\frac{|\langle\psi(\tau)|\hat{a}_1^\dag\hat{b}_1^\dag\hat{a}_2\hat{b}_2|\psi(\tau)\rangle|}{|\langle\psi(0)|\hat{a}_1^\dag\hat{b}_1^\dag\hat{a}_2\hat{b}_2|\psi(0)\rangle|},\nonumber\\
\gamma(\tau)&=&\frac{|\langle\psi(\tau)|\hat{a}_1^\dag\hat{b}_1\hat{a}_2\hat{b}_2^\dag|\psi(\tau)\rangle|}{|\langle\psi(0)|\hat{a}_1^\dag\hat{b}_1\hat{a}_2\hat{b}_2^\dag|\psi(0)\rangle|},
\end{eqnarray}
where the parameters are normalized to the initial values for comparison.

Figure 3 show that the analytical results of mean-field theory (Fig.3(a-d)) agree with the numerical results in the Bose-Hubbard model (Fig.3(e-h)) ($U_{AA}=U_{BB}$). The revival peaks stager for $\beta(\tau)$ and $\gamma(\tau)$ as the inter-species interaction $U_{AB}$ is small. The revival periods of $\beta(\tau)$ and $\gamma(\tau)$ are $T_\beta=2\pi/(U_{AA}+U_{AB})$ and $T_\gamma=2\pi/(U_{AA}-U_{AB})$, respectively. As $U_{AB}/U_{AA}=0.5$, the revival period of $\gamma(\tau)$ is three times of $\beta(\tau)$ (Fig.3(c,g)). Most remarkably, as $U_{AB}=U_{AA}$, $T_\gamma\rightarrow \infty$ implies the counterflow parameter $\gamma(\tau)$ keep invariant throughout the temporal evolution while $\alpha(\tau)$ and $\beta(\tau)$ still experience collapses and revivals (Fig.3(d,h)). Since there are time-intervals of collapses of the single-particle superfluid order parameter $\alpha(\tau)=0$, the non-vanishing (revivals of) pair or anti-pair order parameters $\beta$ or $\gamma$ respectively imply periodical occurrence of CFS or PSF. Another clear indication of PSF or CFS comes from Fig.1(a) at $U_{AB}/U_{AA}=0.1$, where the single-particle SF is nearly damped while the order parameters of the composite operators $\beta$ and $\gamma$ revive periodically.

From Eqs.(\ref{beta}) we note that both repulsive and attractive inter-species interactions can lead to PSF and CFS as well. The observation is clearly distinct with the previous exclaims that PSF requiring $U_{AB}<0$ while CFS requiring $U_{AB}>0$.

\begin{figure}
\begin{center}
\includegraphics*[width=8.5cm]{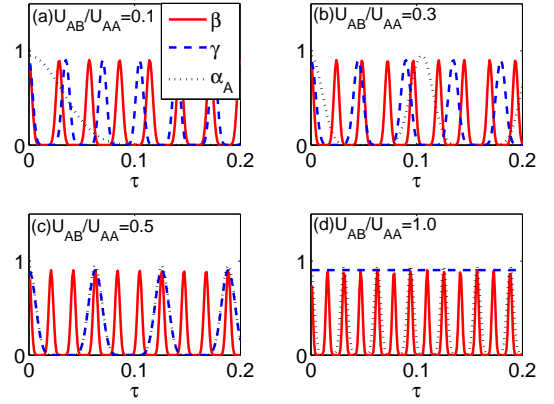}
\caption{(Color online) Dynamical revivals of the coherence $\alpha$, $\beta$, and $\gamma$ evaluated by the independent evolution of the two-site Hubbard model (Eqs.(\ref{independent})) for $U_{BB}/U_{AA}=1.0$. (a) $U_{AB}/U_{AA}=0.1$; (b)
$U_{AB}/U_{AA}=0.3$; (c) $U_{AB}/U_{AA}=0.5$; (d) $U_{AB}/U_{AA}=1.0$}
\end{center}
\end{figure}
It is heuristic to understand the quantum dynamics of the two-species bosons in the double well in the following way. As the system is swiftly changed to the deep MI regimes, the tunneling are suppressed and the delocalized distributions of bosons are frozen. The two-site model degenerate into two individual sites in which the atoms evolve independently. The eigenstates are given by product atom number states of $k$ species-A and $l$ species-B bosons $|\psi_{kl}\rangle=(\hat{a}_j^\dagger)^k(\hat{b}_j^\dagger)^l|0\rangle$, with eigenenergies $E_{kl}=\frac{1}{2}U_{AA}k(k-1)+\frac{1}{2}U_{BB}l(l-1)+U_{AB}kl$. The initial coherent state (\ref{initial}) can be
expanded as
\begin{eqnarray}
|\psi(0)\rangle&=&\frac{1}{2^{(N_A+N_B)/2}}\sum_{k=0}^{N_A}\sum_{l=0}^{N_B}C_{N_A}^kC_{N_B}^l\nonumber\\
&&{\times}(\hat{a}_1^\dag)^{N_A-k}(\hat{a}_2^\dag)^k(\hat{b}_1^\dag)^{N_B-l}(\hat{b}_2^\dag)^l|0\rangle.
\end{eqnarray}
which is a linear superposition of products of the eigenstates $|\psi_{kl}\rangle$. The two sites evolve independently according to
\begin{widetext}
\begin{eqnarray}
|\psi(\tau)\rangle&=&\frac{1}{2^{(N_A+N_B)/2}}\sum_{k=0}^{N_A}\sum_{l=0}^{N_B}\{C_{N_A}^kC_{N_B}^le^{-i[U_{AA}(N_A-k)(N_A-k-1)/2+U_{BB}(N_B-l)(N_B-l-1)/2+U_{AB}(N_A-k)(N_B-l)]\tau}\nonumber\\
&&{\times}(\hat{a}_1^\dag)^{N_A-k}(\hat{b}_1^\dag)^{N_B-l}e^{-i[U_{AA}k(k-1)/2+U_{BB}l(l-1)/2+U_{AB}kl]\tau}(\hat{a}_2^\dag)^k(\hat{b}_2^\dag)^l\}|0\rangle\nonumber\\
&=&\frac{e^{-i\phi(\tau)}}{2^{(N_A+N_B)/2}}\sum_{k=0}^{N_A}\sum_{l=0}^{N_B}C_{N_A}^kC_{N_B}^l\nonumber\\
&&{\times}e^{i[U_{AA}(N_A-k)k+U_{BB}(N_B-l)l+U_{AB}(N_A-k)l+U_{AB}(N_B-l)k]\tau}(\hat{a}_1^\dag)^{N_A-k}(\hat{b}_1^\dag)^{(N_B-l)}(\hat{a}_2^\dag)^k(\hat{b}_2^\dag)^l|0\rangle,
\label{evolution}
\end{eqnarray}
\end{widetext}
where the global phase factor $\phi(\tau)=[U_{AA}N_A(N_A-1)/2+U_{BB}N_B(N_B-1)/2+U_{AB}N_AN_B]\tau$. There is a time-dependent phase factor attached to each term.

The state will evolve back to its initial coherent state after time intervals $T=2\pi/[U_{AA}(N_A-k)k+U_{BB}(N_B-l)l+U_{AB}(N_A-k)l+U_{AB}(N_B-l)k]$. Given $k$ and $l$ being integers, the coherence revivals pose a constraint to the values of the three interactions $U_{AA}$, $U_{BB}$, and $U_{AB}$. It is straightforwardly to evaluate from the definitions,
\begin{widetext}
\begin{eqnarray}
\alpha_A(\tau)&=&|\sum_{k=0}^{N_A}\sum_{l=0}^{N_B}C_{N_A}^{k}C_{N_A}^{k+1}(C_{N_B}^{l})^2e^{2i(U_{AA}k+U_{AB}l)\tau}|/\sum_{k=0}^{N_A}\sum_{l=0}^{N_B}C_{N_A}^{k}C_{N_A}^{k+1}(C_{N_B}^{l})^2,\nonumber\\
\beta(\tau)&=&|\sum_{k=0}^{N_A}\sum_{l=0}^{N_B}C_{N_A}^{k}C_{N_A}^{k+1}C_{N_B}^{l}C_{N_B}^{l+1}e^{2i[(U_{AA}+U_{AB})k+(U_{BB}+U_{AB})l]\tau}|/\sum_{k=0}^{N_A}\sum_{l=0}^{N_B}C_{N_A}^{k}C_{N_A}^{k+1}C_{N_B}^{l}C_{N_B}^{l+1},\nonumber\\
\gamma(\tau)&=&|\sum_{k=0}^{N_A}\sum_{l=0}^{N_B}C_{N_A}^{k}C_{N_A}^{k+1}C_{N_B}^{l}C_{N_B}^{l-1}e^{2i[(U_{AA}-U_{AB})k-(U_{BB}-U_{AB})l]\tau}|/\sum_{k=0}^{N_A}\sum_{l=0}^{N_B}C_{N_A}^{k}C_{N_A}^{k+1}C_{N_B}^{l}C_{N_B}^{l-1}.
\label{independent}
\end{eqnarray}
\end{widetext}
$\alpha_B$ is obtained by substituting ${\hat a}^\dag$ (${\hat a}$) into ${\hat b}^\dag$ (${\hat b}$). The evolution of $\alpha(\tau)$, $\beta(\tau)$ and $\gamma(\tau)$ are displayed in Fig.4. Obviously, they coincide with the exact diagonalization results with non-vanishing hoppings. In the expressions of functions $\alpha(\tau)$, $\beta(\tau)$ and $\gamma(\tau)$ there is a time-dependent phase factor in each term which causes destroy of coherence due to interference. However, after proper time-intervals the phase factors may evolve complete circles of modula $2\pi$ and lead to revivals of coherence. Explicitly, $\alpha_A(\tau)$ will revive to unit after a commensurate period of $T_A=\pi/U_{AA}$ or $T_A=\pi/U_{AB}$. If there is no commensurability between $U_{AA}$ and $U_{AB}$ or $U_{AB}/U_{AA}$ is too small, then the quantum coherence will be damped and no revivals occur (Fig.1(e,f)). If $U_{AB}/U_{AA}=1$, then the revival period is $T_A=\pi/U_{AA}$ (Fig.1(h)). If $U_{AB}/U_{AA}=0.5$, then $T_A=\pi/U_{AB}=2\pi/U_{AA}$ (Fig.1(g)). Similar analyses apply to $\beta(\tau)$ or $\gamma(\tau)$, where the revivals respectively depend on the commensurability between $U_{AA}-U_{AB}$ and $U_{BB}-U_{AB}$ or between $U_{AA}+U_{AB}$ and $U_{BB}+U_{AB}$. In a special case of $U_{AA}=U_{BB}=U_{AB}$, all the phase factors disappear in $\gamma(\tau)$, yielding a constant counterflow coherence (Fig.3(h)).

Experimentally, PSF can be detected by using a Feshbach ramp, similar to what has been used in BEC-BCS experiments . It generates a quasicondensate signal in the resulting molecules. CFS can be detected by applying a $\pi/2$ pulse followed by Bragg spectroscopy which generates a quasicondensate signal in the structure factor. Time-of-flight expansion are used to show the absence of single-particle superfluidity in PSF and CFS.

In summary, we have studied the quantum coherence of two-species bosons within the mean-field theory and exact diagonalization with the two-site Hubbard model. The revival periods are closely related to the interspecies and intraspecies interactions. In the deep Mott regimes ($U_{AA},U_{BB}\gg 1$), we found indications of PSF and CFS in the dynamical evolution at time-intervals of the collapse of the singe-particle SF, where $|\langle\hat{a}\hat{b}(\tau)\rangle|\neq0$ and $|\langle\hat{a}^\dag\hat{b}(\tau)\rangle|\neq0$, while $|\langle\hat{a}\rangle(\tau)|=|\langle\hat{b}\rangle(\tau)|=0$.

This work is supported by funds from the Ministry of Science and Technology of China under Grant
No. 2012CB821403.


\begin{thebibliography}{99}
\bibitem{Paganelli} M. {\L}\c{a}cki, S. Paganelli, V. Ahufinger, A. Sanpera, and J. Zakrzewski, Phys. Rev. A 83, 013605(2011).
\bibitem{Xu} Z. F. Xu, R. L\"{u}, and L. You, Phys. Rev. A 84, 063634 (2011).
\bibitem{Albus} A. Albus, F. Illuminati, and J. Eisert, Phys. Rev. A 68, 023606 (2003).
\bibitem{Imambekov} A. Imambekov, M. Lukin, and E. Demler, Phys. Rev. A 68, 063602 (2003).
\bibitem{Messeguer} B. J. D\'{i}az, M. M. Messeguer, M. Guilleumas, and A. Polls, Phys. Rev. A 80, 043622 (2009).
\bibitem{Chatterjee} B. Chatterjee, I. Brouzos, L. Cao, and P. Schmelcher, Phys. Rev. A 85, 013611 (2012).
\bibitem{Zhou} K. Y. Zhang, L. Zhou, H. Y. Ling, H. Pu, and W. P. Zhang, Phys. Rev. A 83, 063624 (2011).
\bibitem{Tsuchiya} S. Tsuchiya, S. Kurihara, T. Kimura, Phys. Rev. A 70, 043628 (2004).
\bibitem{Sachdev} S. Sachdev, Quantum Phase Teansitions (Cambridge University Press, Cambridge, England, (1999).
\bibitem{Javanainen} J. Javanainen, Phys. Rev. Lett. 57, 3164 (1986).
\bibitem{Anker} Th. Anker, M. Albiez, R. Gati, S. Hunsmann, B. Eiermann, A. Trombettoni, and M. K. Oberthaler, Phys. Rev. Lett. 94, 020403 (2005).
\bibitem{Albiez} M. Albiez, R. Gati, J. F\"{o}lling, S. Hunsmann, M. Cristiani, and M. K. Oberthaler, Phys. Rev. Lett. 95, 010402 (2005).
\bibitem{Winkler} K. Winkler, G. Thalhammer, F. Lang, R. Grimm, J. Hecker Denschlag, A. J. Daley, A. Kantian, H. P. B\"{u}chler, and P. Zoller, Nature (London) 441, 853 (2006).
\bibitem{Will} S. Will, T. Best, S. Braun, U. Schneider, and I. Bloch, Phys. Rev. Lett. 106, 115305 (2011).
\bibitem{Yang} S. J. Yang and S. M. Nie, Phys. Rev. A 82, R061607 (2010).
\bibitem{Zhang} Z. H. Zhang, P. Lu, S. Feng, and S. J. Yang, Phys. Rev. A, 85, 033617 (2012).
\bibitem{Pei} P. Lu, Z. H. Zhang, S. Feng, and S. J. Yang, Phys. Rev. B, 86, 104504 (2012).
\bibitem{Lee1} C. Lee, L.-B. Fu, and Y. S. Kivshar, Euro. Phys. Lett. 81, 60006(2008).
\bibitem{Lee2} C. Lee, Phys. Rev. Lett. 97, 150402 (2006).
\bibitem{Kuklov} A. B. Kuklov and B. V. Svistunov, Phys. Rev. Lett. 90, 100401 (2003).
\bibitem{Demler} E. Demler, F. Zhou, Phys. Rev. Lett. 88, 163001 (2002).
\bibitem{Kagan} M. Y. Kagan, D. V. Efremov, Phys. Rev. B. 65, 195103 (2002).
\bibitem{Svistunov} A. Kuklov, N. Prokof\'ev, B. Svistunov, Phys. Rev. Lett.92, 030403 (2004).
\bibitem{Yang1} S. J. Yang and S. Feng, New J. Phys. 12, 023032 (2010); S. J. Yang, Commun. Theor. Phys. 52, 611 (2009).
\bibitem{Greiner} M. Greiner, O. Mandel, T. W. H\"{a}nsch, and I. Bloch, Nature (London) 419, 51 (2002).
\bibitem{Jaksch} D. Jaksch, C. Bruder, J. I. Cirac, C. W. Gardiner, and P. Zoller, Phys. Rev. Lett. 81, 3108(1998).
\bibitem{Vardi} A. Vardi and J. R. Anglin, Phys. Rev. Lett. 86, 568 (2001).
\bibitem{Krauth} W. Krauth, M. Caffarel, and J.-P. Bouchaud, Phys. Rev. B 45, 3137 (1992).
\bibitem{Qi} Q. Zhou and S. Das Sarma, Phys. Rev. A 82,041601 (2010).
\end{thebibliography}
\end{document}